\documentstyle[aps,prl,epsfig]{revtex}

\newcommand{\tr}{{\rm tr}}

\begin{document}

\title{Calorons in Weyl Gauge}
\author{Gerald V. Dunne}
\address{Department of Physics, University of Connecticut, Storrs, CT
06269-3046, USA}
\author{Bayram Tekin}
\address{Theoretical Physics, University of Oxford, 1 Keble Rd., Oxford,
OX1 3NP, UK}

%\date{\today}
\maketitle
\vskip .5cm

\begin{abstract} 
We demonstrate by explicit construction that while the untwisted
Harrington-Shepard caloron $A_\mu$ is manifestly periodic in
Euclidean time, with period $\beta=\frac{1}{T}$, when transformed to the
Weyl ($A_0=0$) gauge, the caloron gauge field $A_i$ is periodic only
up to a large gauge transformation, with winding number equal to the
caloron's topological charge. This helps clarify the tunneling
interpretation of these solutions, and their relation to Chern-Simons
numbers and winding numbers. 

\end{abstract}

\vskip .5cm
\centerline{PACS: 11.15.-q, 11.10.Wx, 11.27.+d}
\bigskip

%\section{Introduction}
 
Instantons have many important applications in particle physics
\cite{jackiw,coleman,pisarski,shifman,shuryak,thooft}. They may be defined
as classical solutions to the Euclidean Yang-Mills equations that minimize
the Yang-Mills action within a given topological charge sector. Explicit
instanton solutions \cite{polyakov} can be found by solving the
first-order self-duality equation 
\begin{equation}
F_{\mu\nu}=\frac{1}{2}
\epsilon_{\mu\nu\rho\sigma} F_{\rho\sigma}
\label{sd}
\end{equation}
At zero temperature, instantons provide a semiclassical description of
tunneling between classical minima of the Yang-Mills potential that are
connected by large gauge transformations \cite{jr,callan}. At finite
temperature, the simplest instanton solutions are the untwisted
'calorons' of Harrington and Shepard \cite{hs}. These calorons are
solutions to the self-duality equation (\ref{sd}) that are periodic in
Euclidean time, with period
$\beta=\frac{1}{T}$ equal to the inverse temperature. At first sight, it
may seem puzzling that a periodic solution, with
$A_\mu(\vec{x},t=-\frac{\beta}{2})= A_\mu(\vec{x},t=+\frac{\beta}{2})$,
could describe tunneling from one sector to another since the gauge field
$A_\mu$ is the same at either end. The answer is very simple but
illustrative, and does not appear to have been spelled out in detail
before. In the Weyl ($A_0=0$) gauge, which is most convenient for
describing Hamiltonian processes such as tunneling, the coordinate gauge
field $A_i$ is {\it not} periodic; rather,
$A_i(\vec{x},t=+\frac{\beta}{2})$ is related to
$A_i(\vec{x},t=-\frac{\beta}{2})$ by a fixed-time large gauge
transformation whose winding number is equal to the integer topological
charge of the caloron solution. Here we present a simple explicit
construction of this gauge transformation, yielding the Weyl gauge form
of the caloron. We conclude with some comments concerning the distinction
between the caloron solutions and the so-called ``periodic instantons''
studied in \cite{tinyakov,chudnovsky,yaffe,bonini}. For simplicity, we
concentrate in this paper on untwisted calorons \cite{hs}, which are
related to tunneling between sectors of different Chern-Simons charge; we
will not address twisted calorons \cite{kimyeong,vanbaal,vb}, which
involve tunneling between different flux sectors. Twisted calorons are
important for Yang-Mills-Higgs theory with symmetry breaking, and
especially for understanding the monopole content of the theory. The
twisted calorons have nontrivial holonomy, characterized by a nontrivial
Polyakov loop at spatial infinity: ${\cal P} \exp\left(\int_0^\beta dt
A_0\right)$, and the expectation value of $A_0$ at infinity plays the
role of a Higgs expectation value \cite{kimyeong,vanbaal,vb}. Thus, it is
simpler to discuss first the Weyl gauge for untwisted calorons which have
a trivial holonomy.

In discussing SU(2) instantons, it is convenient to use a radially
symmetric ansatz form \cite{witten}. Here, we adopt the conventions of
\cite{jp} for the symmetric ansatz for the SU(2) instanton field
$A_\mu=\frac{\sigma^a}{2i} A^a_\mu$ :
\begin{eqnarray} 
A_i^a&=& {\phi_2-1\over r^2}\,\epsilon_{iak}x_k+\frac{\phi_1}{r}
\left(\delta_{ia}-\frac{x_i x_a}{r^2}\right)+A_1\,\frac{x_i x_a}{r^2}
\nonumber\\ 
A_0^a&=&A_0\, \frac{x_a}{r}
\label{ansatz}
\end{eqnarray} 
where $A_0$, $A_1$, $\phi_1$, and $\phi_2$ are functions of
$r=\sqrt{\vec{x}^2}$ and $t$ only. The charge 1 instanton for SU(2) can
then be expressed in terms of a single function $\rho(r,t)$, with the
ansatz functions in (\ref{ansatz}) given by:
\begin{eqnarray} 
A_0&=&\partial_r \ln \rho\nonumber\\ 
A_1&=&-\partial_0
\ln\rho\nonumber\\
\phi_1&=&-r\,\partial_0 \ln\rho\nonumber\\
\phi_2&=&1+r\partial_r\ln\rho
\label{abelian}
\end{eqnarray} 
Then the gauge field described by (\ref{ansatz},\ref{abelian}) satisfies
the instanton equation (\ref{sd}) if $\rho$ satisfies the linear
equation: $\Box\rho=0$. 
For the zero temperature, charge 1, instanton in SU(2), the
function
$\rho(r,t)$ can be chosen :
\begin{eqnarray}
\rho=1+\frac{\lambda^2}{r^2+t^2}
\label{zerorho}
\end{eqnarray} 
This form corresponds to a single instanton located at the origin
in $R^4$, and with scale parameter $\lambda$. This choice is known as the
`singular
gauge'.

To facilitate the transformation of the gauge field in (\ref{ansatz}) into
the
Weyl ($A_0=0$) gauge, we use the fact \cite{witten} that the ansatz fields
$A_0$, $A_1$, $\phi_1$ and $\phi_2$, which appear in (\ref{ansatz}) and
(\ref{abelian}), transform in an abelian-like manner under a particular
type of
SU(2) gauge transformation on the non-abelian gauge field
$A_\mu=\frac{\sigma^a}{2i} A^a_\mu$. Namely, if the gauge transformation
matrix
$U$ has the special form
\begin{eqnarray} 
U(\vec{x},t)= e^{-\frac{i}{2}f(r,t)\hat{x}\cdot\vec{\sigma}}
=\cos(\frac{f}{2})-i(\hat{x}\cdot\vec{\sigma}) \sin(\frac{f}{2})
\label{trans}
\end{eqnarray} 
then the SU(2) gauge transformation, $A_\mu\to \tilde{A}_\mu=
U^{-1}A_\mu U+U^{-1}\partial_\mu U$, has the following effect on the
ansatz
functions $A_0$, $A_1$, $\phi_1$ and $\phi_2$ :
\begin{eqnarray} 
A_0&\to &\tilde{A_0}=A_0+\partial_0 f \nonumber\\ 
A_1&\to &\tilde{A_1}=A_1+\partial_r f \nonumber\\
\phi_1&\to &\tilde{\phi}_1=\cos f\, \phi_1+\sin f \,\phi_2 \nonumber\\
\phi_2&\to &\tilde{\phi}_2=-\sin f \,\phi_1+\cos f \, \phi_2
\label{abtrans}
\end{eqnarray} 
Thus, to achieve the Weyl gauge, $\tilde{A}_0^a=0$, we simply need
a gauge transformation $U(\vec{x},t)$ of the form in (\ref{trans})
where the function $f(r,t)$ satisfies
\begin{equation}
\partial_0\, f=-\partial_r\,\ln\rho
\label{condition}
\end{equation} 
For the zero temperature, charge 1, instanton with $\rho$ given by
(\ref{zerorho}), this leads to
\begin{eqnarray} 
f(r,t)=\left(\pi+2\arctan\frac{t}{r}\right)-\frac{r}{\sqrt{r^2+\lambda^2}}
\left(\pi+2\arctan\frac{t}{\sqrt{r^2+\lambda^2}}\right)
\label{zerof}
\end{eqnarray} 
where we have chosen the constant of integration so that
$f(r,t=-\infty)=0$. The {\it arctan} function is understood to take values
in the
$[-\frac{\pi}{2},\frac{\pi}{2}]$ branch. The gauge field, $\tilde{A}_i^a$,
in Weyl
gauge has the ansatz form (\ref{ansatz}), with ansatz functions
$\tilde{\phi}_1$, $\tilde{\phi}_2$, $\tilde{A}_1$, (and $\tilde{A}_0\equiv
0$)
given by (\ref{abtrans}).

The tunneling interpretation of the instanton solution becomes evident by
noting
that $\tilde{A}_i$ is pure gauge at $t=\pm\infty$. At $t=-\infty$, 
\begin{eqnarray}
\tilde{A}_i^a(\vec{x},t=-\infty)=0
\label{minuszero}
\end{eqnarray} while at $t=+\infty$,
\begin{eqnarray}
\tilde{A}_i^a(\vec{x},t=+\infty)&=&{\cos\left(\frac{2\pi
r}{\sqrt{r^2+\lambda^2}}\right)-1\over r^2}\,\epsilon_{iak} x_k- 
{\sin\left(\frac{2\pi r}{\sqrt{r^2+\lambda^2}}\right)\over r}
\left(\delta_{ia}-\frac{x_ix_a}{r^2}\right)  - 
{2\pi\lambda^2\over (r^2+\lambda^2)^{3/2}}\,\frac{x_ix_a}{r^2}
\nonumber\\
&=&\left(W^{-1}\partial_i W\right)^a
\label{pluszero}
\end{eqnarray} 
where the static gauge function $W(\vec{x})$ is
\begin{eqnarray} 
W(\vec{x})=e^{-\frac{i}{2}f(r,t=+\infty)
\hat{x}\cdot\vec{\sigma}}= -\exp\left[i\frac{\pi r}
{\sqrt{r^2+\lambda^2}}\,
\hat{x}\cdot\vec{\sigma}\right]
\label{uzero}
\end{eqnarray} 
Thus, as $t$ goes from $t=-\infty$ to $t=+\infty$, the instanton
gauge field $\tilde{A}_i$ in Weyl gauge interpolates between the pure
gauge $\tilde{A}_i=0$ in (\ref{minuszero}), and the pure gauge
$\tilde{A}_i=W^{-1}\partial_i W$ in (\ref{pluszero}). These fields
$\tilde{A}_i$ are neighboring minima (in fact, zeros) of the classical
Yang-Mills potential
\begin{eqnarray} 
V=\frac{1}{2}\int d^3 x\, B^a_i B^a_i
\label{pot}
\end{eqnarray}
>From (\ref{minuszero}) and (\ref{pluszero}) we see that these two
minima are related by the large gauge transformation $W(\vec{x})$ in
(\ref{uzero}). Furthermore, the fixed-time gauge transformation in
(\ref{uzero}) is
$W(\vec{x})=U(\vec{x},t=+\infty)$, where $U(\vec{x},t)$ is the
(time-dependent) gauge transformation used to transform to the Weyl
gauge. To evaluate the winding
number of the fixed-time gauge transformation $W$, we note that if a
gauge transformation has the form
$W(\vec{x})=e^{-\frac{i}{2}g(r)\hat{x}\cdot\vec{\sigma}}$, then the
winding number $N$ is determined \cite{roman} by the values of $g(r)$ at
$r=0$ and $r=\infty$ : 
\begin{eqnarray}
N\equiv \frac{1}{24\pi^2}\int\, d^3x\, \epsilon_{ijk}\,
\tr\left(W^{-1}\partial_i W W^{-1}\partial_j W W^{-1}\partial_k W\right)=
-\frac{1}{2\pi}\left[g(r)-\sin g(r)\, \right]^{r=\infty}_{r=0}
\label{winding}
\end{eqnarray}
Thus, for the gauge transformation in (\ref{uzero}), we find $N=1$.

Having reviewed this familiar tunneling interpretation of zero temperature
instantons, we now turn to finite temperature instantons, or `calorons'.
At finite
T, the Euclidean time direction is compactified onto the finite interval
$t\in
[-\frac{\beta}{2},+\frac{\beta}{2}]$, where
$\beta=\frac{1}{T}$ is the inverse temperature. The manifold is now
$R^3\times S^1$, rather than $R^4$. Instantons are solutions to the same
self-duality equation (\ref{sd}), but the gauge fields, being bosonic,
satisfy periodic boundary conditions \cite{claude,pisarski} on the
Euclidean time interval $[-\frac{\beta}{2},+\frac{\beta}{2}]$.

The untwisted Harrington-Shepard (HS) caloron solution is consistent with
the
ansatz form described in (\ref{ansatz}) and (\ref{abelian}). The only
change is
that the function $\rho(r,t)$ becomes \cite{hs}
\begin{eqnarray}
\rho(r,t)&=&1+\frac{1}{2}\left(\frac{2\pi \lambda}{\beta}\right)^2
\left(\frac{\beta}{2\pi r}\right){\sinh\left(\frac{2\pi r}{\beta}\right)
\over \cosh\left(\frac{2\pi r}{\beta}\right) -
\cos\left(\frac{2\pi t}{\beta}\right)}\nonumber\\
&\equiv&1+\frac{1}{2}\frac{\bar{\lambda}^2}{\bar{r}}{\sinh\bar{r}\over
\cosh\bar{r}-\cos\bar{t}}
\label{finiterho}
\end{eqnarray} 
where we have defined the convenient short-hand: $\bar{r}=\frac{2\pi
r}{\beta}$, etc. Note that $\rho$ in (\ref{finiterho}) satisfies
$\Box\rho=0$. The caloron solution given by (\ref{ansatz}),
(\ref{abelian}) and (\ref{finiterho}) corresponds to an infinite line of
zero temperature instantons, each of scale $\lambda$ and each at spatial
position
$\vec{x}=0$, lined up along the t axis, with period $\beta$. It is 
straightforward to check that in the zero temperature limit
($\beta\to\infty$), the expression for
$\rho(r,t)$ in (\ref{finiterho}) reduces smoothly to that in
(\ref{zerorho}), and we recover the zero temperature instanton. At finite
T, the caloron solution (\ref{ansatz})-(\ref{abelian}) is manifestly
periodic in $t$,
$A_\mu(\vec{x},t+\beta)=A_\mu(\vec{x},t)$, since $\rho(r,t)$ in
(\ref{finiterho})
has period $\beta$.

To explore the tunneling interpretation of this caloron solution, we
transform it to the Weyl gauge. Once again, the Weyl gauge can be attained
most easily by using the abelian-like transformation properties
(\ref{abtrans}) of the ansatz functions in (\ref{ansatz}). The required
SU(2) transformation matrix has the form in (\ref{trans}), with $f(r,t)$
satisfying (\ref{condition}), and $\rho$ is now given by
(\ref{finiterho}). The solution to (\ref{condition}) in this case is
\begin{eqnarray}
f(r,t)&=&\left(\pi+2\arctan\left[\coth(\frac{\bar{r}}{2})\tan(\frac{\bar{t}}{2})
\right]\right)+\nonumber\\
&&\hskip-2cm{(\bar{\lambda}^2-2\bar{r}^2)\sinh\bar{r}-\bar{\lambda}^2\bar{r}
\cosh\bar{r}\over \bar{r}
\sqrt{\sinh\bar{r}(4\bar{\lambda}^2\bar{r}\cosh\bar{r}
+(\bar{\lambda}^4+4\bar{r}^2)\sinh\bar{r})}} \left(\pi+2\arctan\left[
\tan(\frac{\bar{t}}{2}){\bar{\lambda}^2\sinh\bar{r}+2\bar{r}(1+\cosh\bar{r})
\over
\sqrt{\sinh\bar{r}(4\bar{\lambda}^2\bar{r}\cosh\bar{r}
+(\bar{\lambda}^4+4\bar{r}^2)\sinh\bar{r})}}\right]\right)
\label{finitef}
\end{eqnarray} 
where the constant of integration has been chosen so that
$f(r,t=-\frac{\beta}{2})=0$.
Note that this function $f(r,t)$, and hence the gauge
transformation
$U(\vec{x},t)=\exp(-\frac{i}{2}f(r,t)\,\hat{x}\cdot\vec{\sigma})$ that
transforms the HS caloron to the Weyl gauge, is manifestly periodic in t,
with period $\beta$.  It is straightforward to check that $f(r,t)$ in
(\ref{finitef}) reduces smoothly, as $\beta\to\infty$, to the zero
temperature function in (\ref{zerof}).

Given the function $f(r,t)$ in (\ref{finitef}), the caloron fields
$\tilde{A}_i^a$ in the Weyl gauge are obtained simply from (\ref{ansatz})
and
(\ref{abtrans}). To investigate the relation of this caloron solution to
quantum
tunneling, we  evaluate $\tilde{A}_i^a$ at
$t=\pm\frac{\beta}{2}$. At
$t=-\frac{\beta}{2}$, the ansatz fields satisfy
$\tilde{A}_1=\tilde{\phi}_1=0$, while
$\tilde{\phi}_2=[1+r\partial_r\ln \rho(r,t=-\frac{\beta}{2})]$,
where
$\rho(r,t=-\frac{\beta}{2})=1+\frac{\bar{\lambda}^2}{2\bar{r}}
\tanh(\frac{\bar{r}}{2})$. Thus, 
\begin{eqnarray}
\tilde{A}_i^a(\vec{x},t=-\frac{\beta}{2})= -
{\lambda^2\pi\left(\tanh(\frac{\pi
r}{\beta})-\frac{\pi r}{\beta} {\rm sech}^2(\frac{\pi
r}{\beta})\right)\over 
\beta r^3\left(1+\frac{\lambda^2\pi}{r\beta}\tanh(\frac{\pi
r}{\beta})\right)}
\,\epsilon_{iak}x_k 
\label{gaugeminus}
\end{eqnarray} 
For low temperatures,
\begin{eqnarray}
\tilde{A}_i^a (\vec{x},t=-\frac{\beta}{2}) \sim -
\frac{2}{3}\lambda^2\pi^4 T^4\epsilon_{iak}x_k 
\label{lowlimit}
\end{eqnarray} 
which is nonzero, but which vanishes at zero temperature, as in
(\ref{minuszero}). At $t=+\frac{\beta}{2}$, the expression for
$\tilde{A}_i^a(\vec{x},t=+\frac{\beta}{2})$ is more cumbersome to write
out, but it is completely specified by the ansatz form (\ref{ansatz}) with
\begin{eqnarray}
\tilde{\phi}_1(r,t=+\frac{\beta}{2})&=&\sin f(r,t=+\frac{\beta}{2})\,
\phi_2(r,t=+\frac{\beta}{2})\nonumber\\
\tilde{\phi}_2(r,t=+\frac{\beta}{2})&=&\cos f(r,t=+\frac{\beta}{2})\,
\phi_2(r,t=+\frac{\beta}{2})\nonumber\\
\tilde{A}_1(r,t=+\frac{\beta}{2})&=&\partial_r f(r,t=+\frac{\beta}{2})
\label{finiteplus}
\end{eqnarray} 
and where $f$ is given by (\ref{finitef}), and $\phi_2$ is as in
(\ref{abelian}).

We now ask how $\tilde{A}_i^a(\vec{x},t=+\frac{\beta}{2})$ is related to
$\tilde{A}_i^a(\vec{x},t=-\frac{\beta}{2})$. Since the ansatz functions
satisfy $\tilde{A}_1(r,t=-\frac{\beta}{2})=0$, and
$\phi_2(r,t=+\frac{\beta}{2})=\tilde{\phi}_2(r,t=-\frac{\beta}{2})$, this
shows
that $\tilde{A}_i^a(\vec{x},t=+\frac{\beta}{2})$ and
$\tilde{A}_i^a(\vec{x},t=-\frac{\beta}{2})$ are related by a gauge
transformation
\begin{eqnarray}
\tilde{A}_i(\vec{x},t=+\frac{\beta}{2})=W^{-1}\,
\tilde{A}_i(\vec{x},t=-\frac{\beta}{2})\, W+W^{-1}\partial_i W
\label{finitetrans}
\end{eqnarray} 
where the static large gauge transformation $W(\vec{x})$ is
\begin{eqnarray}
W(\vec{x})&=&e^{-\frac{i}{2}f(r,t=+\frac{\beta}{2})\hat{x}\cdot\vec{\sigma}}
\nonumber\\
&=&-\exp\left[i\pi{(\bar{\lambda}^2-2\bar{r}^2)\sinh\bar{r}-\bar{\lambda}^2\bar{r}
\cosh\bar{r}\over
\bar{r}\sqrt{\sinh\bar{r}(4\bar{\lambda}^2\bar{r}\cosh\bar{r}
+(\bar{\lambda}^4+4\bar{r}^2)\sinh\bar{r})}}\,
(\hat{x}\cdot\vec{\sigma})\right]
\label{finiteu}
\end{eqnarray} 
Using (\ref{winding}), the winding number of this gauge transformation is
\begin{eqnarray}
N&=&-\frac{1}{2\pi}\left[f(r,t=+\frac{\beta}{2})-\sin
f(r,t=+\frac{\beta}{2})\right]^{r=\infty}_{r=0}\nonumber\\ &=&1
\label{finitewinding}
\end{eqnarray} 
for {\it all} inverse temperatures $\beta$ and scales $\lambda$.

Notice that (\ref{finitetrans}) shows that in the Weyl gauge the gauge
fields
$\tilde{A}_i$ are {\it not} strictly periodic. Rather,
$\tilde{A}_i$ is periodic up to a large gauge transformation. This
fixed-time gauge transformation is 
$W(\vec{x})=U(\vec{x},t=+\frac{\beta}{2})$, where $U(\vec{x},t)$ is the
time-dependent gauge transformation used to transform to the Weyl
gauge. So, just as in the
$T=0$ case, the initial and final gauge fields,
$\tilde{A}_i(\vec{x},t=\mp\frac{\beta}{2})$, are related by a static
large gauge transformation of unit winding number. But at finite T, these
initial and final gauge fields
$\tilde{A}_i(\vec{x},t=\mp\frac{\beta}{2})$ are not pure gauge. This
means that the tunneling at finite T is not between {\it minima} of the
classical Yang-Mills potential (\ref{pot}), but between fields of higher
potential energy. With the ansatz (\ref{ansatz}), the
corresponding magnetic field strength is (and since the
instanton solution is self-dual, this also gives the electric field
strength)
\begin{eqnarray} 
B_i^a&=&-{\left(\partial_r \phi_1-A_1 \phi_2\right)\over r^2}
\epsilon_{iak} x_k +{\left(\partial_r \phi_2+A_1 \phi_1\right)\over r}
\left(\delta_{ia} -\frac{x_i x_a}{r^2}\right)
+{\left(1-\phi_1^2-\phi_2^2\right)\over r^4} x_i x_a
\label{mag}
\end{eqnarray}  
It is straightforward to verify that
$\tilde{B}_i^a(\vec{x},t)$ is nonvanishing at
$t=\mp\frac{\beta}{2}$. The magnetic potential
energy (\ref{pot}) can be expressed in terms of the ansatz functions in
(\ref{ansatz}) and (\ref{abelian}) [since $V$ is gauge invariant, it
doesn't matter whether we use the original gauge or the Weyl gauge] :
\begin{eqnarray}
V=2\pi \int_0^\infty dr\left\{2(\partial_r \phi_1-A_1
\phi_2)^2+2(\partial_r \phi_2+A_1 \phi_1)^2+\frac{1}{r^2} (1-\phi_1^2
-\phi_2^2)^2\right\}
\label{potansatz}
\end{eqnarray}
At zero temperature, $V(t=\mp\infty)=0$, since the zero
temperature instanton is a pure gauge at $t=\mp\infty$. But at
finite temperature, $V(t=\mp\frac{\beta}{2})$ is nonzero, even though it
approaches zero as $\beta\to \infty$. It is straightforward to evaluate
numerically $V$ in (\ref{potansatz}) for the zero and finite temperature
instantons discussed above. In Figure 1 we plot the potential energy
as a function of Euclidean time for the zero temperature instanton for
which
$\rho(r,t)$ is given by (\ref{zerorho}), and in Figure 2 for the
finite temperature caloron for which $\rho(r,t)$ is given by
(\ref{finiterho}). 

\begin{figure}[htb]
\centering{\epsfig{file=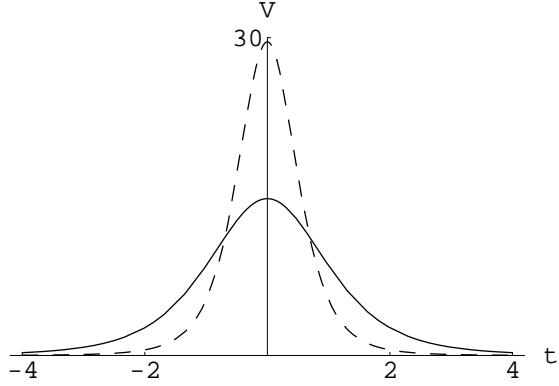}}
\caption{The potential energy (\protect{\ref{potansatz}}) as a function of
$t$, at zero temperature. The dashed line is for instanton scale
parameter $\lambda=1$, while the solid line has $\lambda=2$. Note that
$V\to 0$ at $t=\pm\infty$, as the Weyl gauge  solutions are pure gauge at
$t=\pm\infty$.}
\end{figure}

\begin{figure}[htb]
\centering{\epsfig{file=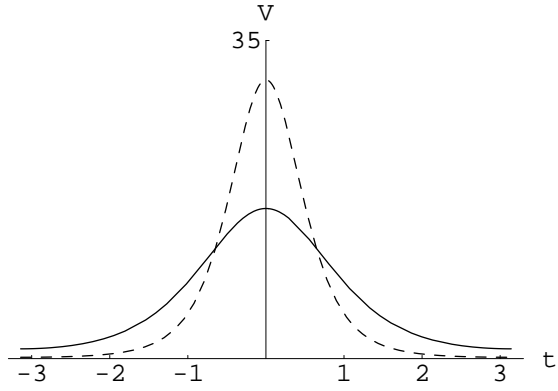}}
\caption{The potential energy (\protect{\ref{potansatz}}) as a function of
$\bar{t}=\frac{2\pi t}{\beta}$, at finite temperature. The dashed
line is for scale parameter
$\bar{\lambda}=1$, while the solid line is for $\bar{\lambda}=2$. Note
that $V$ does not vanish at $\bar{t}=\pm\pi$, as the Weyl gauge solutions
are not pure gauge at $t=\pm\frac{\beta}{2}$.}
\end{figure}

We now consider the instanton's topological charge and its relation with
Chern-Simons numbers and winding numbers. Recall that the instanton charge
may be expressed as
\begin{eqnarray}
Q&=&-{1\over 32 \pi^2}\int d^4 x\, \epsilon_{\mu\nu\rho\sigma}{\rm
tr}\left( F_{\mu\nu} F_{\rho\sigma}\right) \nonumber\\
&\equiv &\int d^4x \, \partial_\mu K_\mu ,\quad {\rm where}\quad  
K_\mu=-\frac{1}{8\pi^2}\epsilon_{\mu\nu\rho\sigma} {\rm tr}\left(
A_\nu \partial_\rho A_\sigma +\frac{2}{3} A_\nu A_\rho A_\sigma\right)
\end{eqnarray}
This leads to a natural separation of Q as
\begin{eqnarray}
Q=\int dt \frac{d}{dt} CS(t) +\int_0^\infty dr \frac{d}{dr} K(r)
\label{split}
\end{eqnarray}
In terms of the ansatz fields in (\ref{ansatz}), the Chern-Simons term,
$CS\equiv \int d^3x K_0$, is
\begin{eqnarray}
CS(t)= \frac{1}{2\pi} \int_0^\infty dr\left[\phi_1 \phi_2^\prime -\phi_2
\phi_1^\prime -A_1(1-\phi_1^2-\phi_2^2) +\phi_1^\prime\right]
\label{cs}
\end{eqnarray}
while 
\begin{eqnarray}
K(r)=\frac{1}{2\pi} \int dt\left[-\phi_1 \dot{\phi_2}
+\phi_2\dot{\phi_1} +A_0(1-\phi_1^2-\phi_2^2) -\dot{\phi_1}\right]
\label{kk}
\end{eqnarray}
Here $\phi^\prime$ and  $\dot{\phi}$ denote derivative with respect to $r$
and $t$ respectively.
At zero temperature, the $t$ integration in (\ref{kk}) is from
$-\infty$ to $+\infty$, while at finite temperature it is over
$[-\frac{\beta}{2},+\frac{\beta}{2}]$. Note that both
$CS(t)$ and $K(r)$ are gauge dependent, while
$Q$ itself is gauge invariant. Also note that the last terms in each of
$CS(t)$ and
$K(r)$ cancel one another in the decomposition (\ref{split}), but are
required for the correct transformation properties of the respective
terms. For example, under a gauge transformation of the form
(\ref{trans}), the Chern-Simons term changes as
\begin{eqnarray}
\widetilde{CS}&=&CS+\frac{1}{8\pi^2}\int d^3x\, \epsilon_{ijk} \partial_i
{\rm tr} \left(\partial_j U U^{-1} A_k\right)+\frac{1}{24\pi^2} \int d^3x
\,\epsilon_{ijk} {\rm tr}\left(U^{-1}\partial_i U U^{-1}\partial_j U
U^{-1}\partial_k U\right)\nonumber\\
&=& CS+\frac{1}{2\pi} \int_0^\infty dr\frac{d}{dr}\left[(\phi_2-1) \sin
f +\phi_1 (\cos f-1)\right] -\frac{1}{2\pi}  \int_0^\infty
dr\frac{d}{dr}\left[f-\sin f\right]
\label{change}
\end{eqnarray}
It is a simple but instructive exercise to check that this nonabelian
transformation law agrees with the transformation of (\ref{cs}) under the
abelian-like transformations (\ref{abtrans}) of the ansatz functions,
provided one includes the $\phi_1^\prime$ term in $CS(t)$ \cite{yaffe2}. 

For the instanton solutions studied in this paper, one can evaluate
numerically the functions $CS(t)$ and $K(r)$ in (\ref{cs}) and
(\ref{kk}), in the original gauge (\ref{abelian}), and in the Weyl gauge
$\tilde{A}_i$ where
$\tilde{A}_0=0$. At zero temperature, with the scalar function
$\rho(r,t)$ given by (\ref{zerorho}), the functions $CS(t)$ and $K(r)$
are plotted in Figures 3 and 4. In this gauge,
$CS(t=+\infty)-CS(t=-\infty)=0$, and
$K(r=\infty)-K(r=0)=1$. Thus, in this gauge the entire contribution to the
instanton charge $Q$ comes from the $K(r)$ term in the
decomposition (\ref{split}). On the other hand, in the Weyl gauge, the
functions $\widetilde{CS}(t)$ and $\tilde{K}(r)$ are plotted in Figures 5
and 6. In the Weyl gauge, the entire contribution to $Q$ comes from the
change in the Chern-Simons number. Also note that in the Weyl gauge the
Chern-Simons number is an integer at $t=\pm\infty$; this is because at
$t=\pm\infty$ the Weyl gauge field is pure gauge [see (\ref{minuszero})
and (\ref{pluszero})], and for a pure gauge field the Chern-Simons term is
equal to the winding number of the corresponding group element. The
time-dependent gauge transformation
$U=\exp(-\frac{i}{2}f(r,t)\,\hat{x}\cdot\vec{\sigma})$, with $f(r,t)$ in
(\ref{zerof}), that transforms from the original singular gauge to the
Weyl gauge can be considered to have a t-dependent winding number $N(t)$
that, from (\ref{winding}) and (\ref{zerof}), is a step function in time:
\begin{eqnarray}
N(t)=\frac{1}{2}\left(1+{\rm sign}(t)\right)
\label{step}
\end{eqnarray}
This is consistent with the plots in Figures 3 and 5, since the middle
term in (\ref{change}) vanishes in this case.
Thus, in
the Weyl gauge, the Chern-Simons number provides a meaningful label for
the distinct classical wells of the Yang-Mills potential. At $t=0$, the
Weyl gauge Chern-Simons number is $\frac{1}{2}$, corresponding to the
sphaleron at the peak of the barrier between neighboring wells. The
winding number of the gauge transformation $U(\vec{x},t)$ is also
$\frac{1}{2}$ when $t=0$.

\begin{figure}[htb]
\centering{\epsfig{file=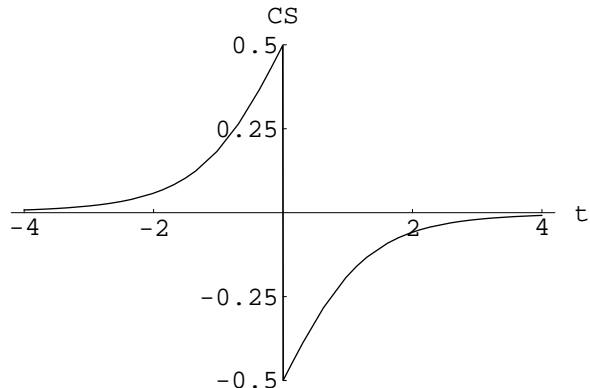}}
\caption{Plot of the Chern-Simons function $CS(t)$ in
(\protect{\ref{cs}}), in the singular gauge at zero temperature, as a
function of $t$. The instanton scale factor $\lambda$ has been scaled to
1.}
\end{figure}

\begin{figure}[htb]
\centering{\epsfig{file=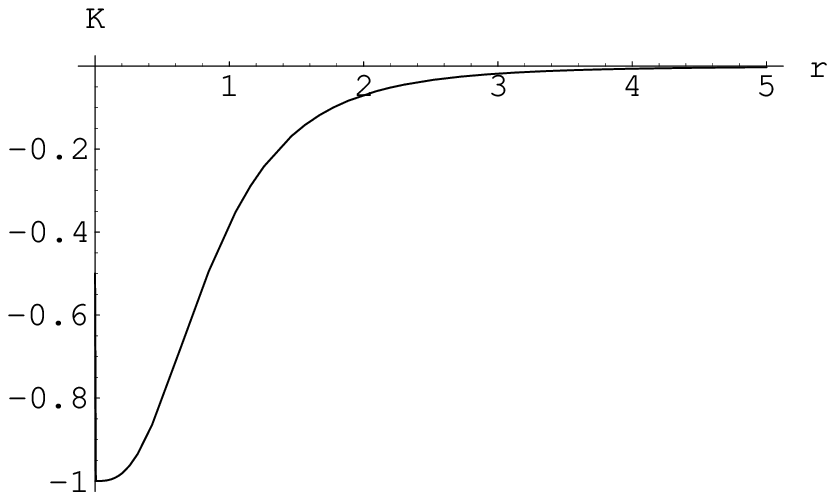}}
\caption{Plot of the function $K(r)$ in (\protect{\ref{kk}}), in the
singular gauge at zero temperature, as a function of $r$. The
instanton scale factor $\lambda$ has been scaled to 1.}
\end{figure}

\begin{figure}[htb]
\centering{\epsfig{file=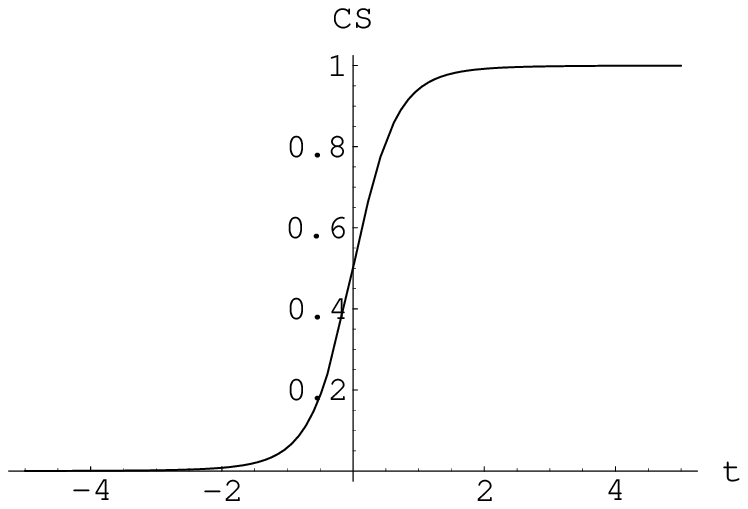}}
\caption{Plot of the Chern-Simons function $CS(t)$
in (\protect{\ref{cs}}), in the Weyl gauge at zero temperature, as a
function of $t$. The instanton scale factor $\lambda$ has been scaled to
1.}
\end{figure}

\begin{figure}[htb]
\centering{\epsfig{file=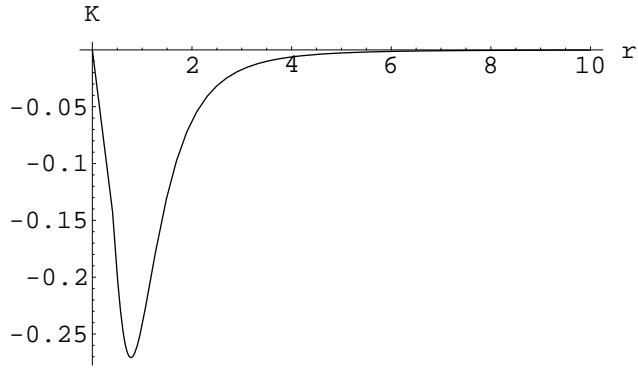}}
\caption{Plot of the function $K(r)$ in (\protect{\ref{kk}}), in the
Weyl gauge at zero temperature, as a function of $r$. The
instanton scale factor $\lambda$ has been scaled to 1.}
\end{figure}

At finite temperature, the situation is similar. The only change is
that the scalar function
$\rho(r,t)$ is now given by (\ref{finiterho}), and the t integration in
(\ref{kk}) is over the finite interval
$[-\frac{\beta}{2},+\frac{\beta}{2}]$. In the Harrington-Shepard gauge
(\ref{abelian},\ref{finiterho}), the functions
$CS(t)$ and $K(r)$ are plotted in Figures 7 and 8. (These plots, and
those in Figs. 9 and 10, are for scale parameter $\bar{\lambda}=1$. For
different $\bar{\lambda}$, the plots have the same form.)  Once
again, in this gauge, the entire contribution to the instanton charge
comes from the change in
$K(r)$. The corresponding plots for the Weyl gauge are in Figures 9 and
10, and we see that the entire contribution to
$Q$ comes from the change
$CS(t)$. From the fact (\ref{finitetrans}) that in the Weyl gauge
$\tilde{A}_i(t=+\frac{\beta}{2})$ is related to
$\tilde{A}_i(t=-\frac{\beta}{2})$ by a static gauge transformation of
winding number 1, we know that the difference
$\widetilde{CS}(t=+\frac{\beta}{2})-\widetilde{CS}(t=-\frac{\beta}{2})=1$.
But it is interesting to note further that in the Weyl gauge the
Chern-Simons number is an integer at $t=\pm \frac{\beta}{2}$, even though
the gauge field is {\it not} a pure gauge [see (\ref{gaugeminus}) and
(\ref{finitetrans})]. At $t=0$, the Weyl gauge Chern-Simons number is
$\frac{1}{2}$, as is the winding number of the gauge transformation
$U(\vec{x},t)$ that connects the HS gauge to the Weyl gauge.

\begin{figure}[htb]
\centering{\epsfig{file=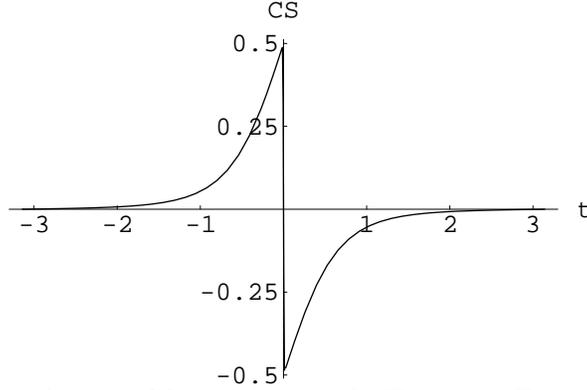}}
\caption{Plot of the Chern-Simons function $CS(t)$ in
(\protect{\ref{cs}}), in the Harrington-Shepard gauge at finite
temperature, as a function of $\bar{t}=\frac{2\pi t}{\beta}$. The scale
factor has been chosen $\bar{\lambda}=1$.}
\end{figure}

\begin{figure}[htb]
\centering{\epsfig{file=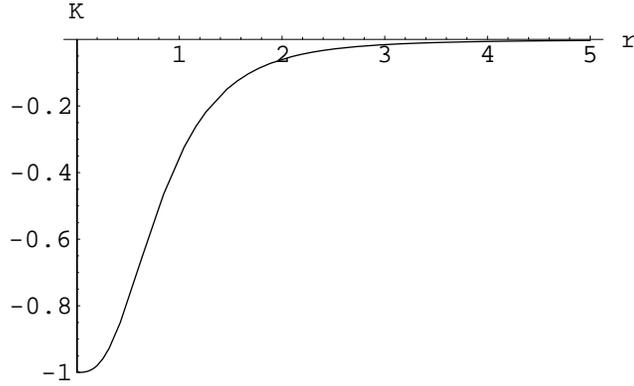}}
\caption{Plot of the function $K(r)$ in (\protect{\ref{kk}}), in the
Harrington-Shepard gauge at finite temperature, as a function of
$\bar{r}=\frac{2\pi r}{\beta}$. The scale  factor has been chosen
$\bar{\lambda}=1$.}
\end{figure}

\begin{figure}[htb]
\centering{\epsfig{file=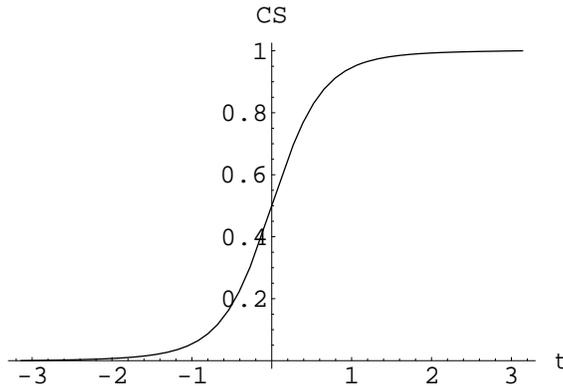}}
\caption{Plot of the Chern-Simons function $CS(t)$ in
(\protect{\ref{cs}}), in the Weyl gauge at finite temperature, as a
function of $\bar{t}=\frac{2\pi t}{\beta}$. The scale factor has been
chosen $\bar{\lambda}=1$.}
\end{figure}

\begin{figure}[htb]
\centering{\epsfig{file=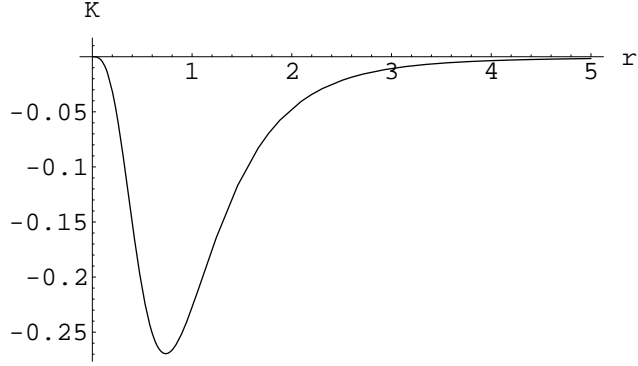}}
\caption{Plot of the function $K(r)$ in (\protect{\ref{kk}}), in the
Weyl gauge at finite temperature, as a function of
$\bar{r}=\frac{2\pi r}{\beta}$. The scale  factor has been chosen
$\bar{\lambda}=1$.}
\end{figure}

In conclusion, we have demonstrated explicitly that while the HS caloron
solution
is manifestly periodic, so that 
$A_\mu(\vec{x},t=+\frac{\beta}{2})=A_\mu(\vec{x},t=-\frac{\beta}{2})$,
when transformed to the Weyl ($A_0=0$) gauge, the gauge field is periodic
up to a large gauge transformation,
$\tilde{A}_i(\vec{x},t=+\frac{\beta}{2})=W^{-1}
\tilde{A}_i(\vec{x},t=-\frac{\beta}{2}) W+W^{-1}\partial_i W$, where $W$
is a fixed-time gauge transformation with integer winding number equal to
the caloron's topological charge. This gauge transformation $W$ connects
different Chern-Simons-number sectors of the classical Yang-Mills
potential energy, just as is familiar from zero temperature. However, the
caloron solution in Weyl gauge interpolates, as $t$ goes from
$-\frac{\beta}{2}$ to $+\frac{\beta}{2}$, not between classical
minima but between classical gauge fields of {\it nonzero} potential
energy. 

Having understood the tunneling interpretation of the HS calorons in Weyl
gauge, we conclude with some brief comments concerning the difference
between these calorons and the "periodic instantons" studied in
\cite{tinyakov,chudnovsky,yaffe,bonini}. These are very different
objects, but the names can be somewhat confusing since calorons are
themselves instantons that are periodic in Euclidean time. This
distinction, and some of its physical consequences, have been stated
clearly in \cite{tinyakov}, but we re-iterate these distinctions here in
the light of our Weyl gauge calorons. The "periodic instantons" are most
easily introduced in the simple context of a quantum mechanical model; for
example, a double well potential $V=(q^2-1)^2/4$. This captures most of
the essential physics. This potential model has well-known instanton
solutions \cite{harrington} that are periodic in Euclidean time t: 
\begin{eqnarray}
q(t)=\sqrt{\frac{2 \nu}{1+\nu}}\,\, {\rm sn}\left(
\frac{t}{\sqrt{1+\nu}}-K\,;\nu\right),
\label{dw}
\end{eqnarray}
where ${\rm sn}(x;\nu)$ is a Jacobi elliptic function \cite{jacobi}, and
the modulus parameter $0\leq\nu\leq 1$ is related to the energy by
$E=(\nu-1)^2/(4(\nu+1)^2)$. These solutions are manifestly periodic:
$q(t=-\frac{\beta}{2})=q(t=+\frac{\beta}{2})$, where the period
$\beta=4\sqrt{1+\nu}\, K(\nu)$, and
$K(\nu)$ is the elliptic quarter-period \cite{jacobi}. Physically, these
periodic instantons correspond to tunneling from one well to the next,
and then back again to the original starting position (since they are
periodic!). By contrast, the Weyl gauge calorons describe tunneling from
one classical well (characterized by Chern-Simons number 0) to the next
well (with CS=1); they do not tunnel back again, since the Weyl gauge
field $\tilde{A}_i(t=+\frac{\beta}{2})$ differs from
$\tilde{A}_i(t=-\frac{\beta}{2})$ by a large gauge transformation, as we
have demonstrated explicitly in (\ref{finitetrans},\ref{finiteu}). This
difference between calorons and periodic instantons also manifests itself
through the fact that the calorons have nonzero topological charge per
period (the explicit ones studied here have Q=1), while the "periodic
instantons" have zero net topological charge per period. For low
temperatures, the periodic instantons can be approximated by an infinite
chain of alternating zero temperature instantons and anti-instantons.
Indeed, this is particularly easy to see in the quantum mechanical
double-well potential case (\ref{dw}), owing to the remarkable elliptic
function identity:
\begin{eqnarray}
{\rm sn}\left(
\frac{t}{\sqrt{1+\nu}}\,;\nu\right)=\frac{\pi}{2\sqrt{\nu}K'}\,
\sum_{n=-\infty}^\infty (-1)^n\, \tanh\left(\frac{\pi}{2\sqrt{1+\nu}
K'} (t-n \frac{\beta}{2})\right)
\end{eqnarray}
Here $K'(\nu)\equiv K(1-\nu)$. This leads to the simple interpretation of
the periodic instanton in (\ref{dw}) as a series of alternating instantons
and anti-instantons. The zero temperature limit corresponds to
$\nu\to 1$, in which case the period $\beta\to \infty$,
and $E\to 0$, and we approach the usual zero temperature kink-like
instantons. But, on the full period
$[-\frac{\beta}{2},\frac{\beta}{2}]$, the $sn$ function reduces to a
{\it pair} of an anti-kink and a kink. This explains why the
net topological charge per period is zero. It is only on the half-period
$[0,\frac{\beta}{2}]$ that $q(t)$ approaches the familiar zero
temperature instanton: $q=\tanh(t/\sqrt{2})$. By contrast, the HS caloron
is a periodic sum of instantons (and an HS anti-caloron is a periodic sum
of anti-instantons), {\it not} a periodic sum of alternating instantons
and anti-instantons. Indeed, in Yang-Mills theory there is no known exact
"periodic instanton" that is an exact alternating chain of instantons and
anti-instantons, although approximate chains of this form have been
extensively studied \cite{tinyakov,shuryak}. This difference between
calorons and periodic instantons can be reconciled because even though
the  caloron solutions are manifestly periodic, 
$A_\mu(t+\beta)=A_\mu(t)$, since this is a gauge theory, the
$A_\mu$ are not all independent physical fields; when we transform
the caloron to the Weyl gauge we see that the coordinate fields
$\tilde{A}_i$ are periodic only up to a large gauge tranformation. We
hope that our explicit construction of the Weyl gauge caloron fields
helps to clarify this important distinction between calorons and
"periodic instantons".

\vskip .1cm {\bf Acknowledgments:}

GD is supported by the U.S. DOE grant DE-FG02-92ER40716.00. GD also thanks
Balliol College, Oxford, the Theoretical Physics Department at Oxford,
and the CSSM at Adelaide University, where this work was begun. BT is
supported by PPARC.

%\end{thebibliography}

\end{document}